\documentclass[aps,twocolumn,prb,tightenlines,floatfix,showpacs]{revtex4}
\usepackage[dvips]{graphicx}
\usepackage[english]{babel}
\usepackage{amsmath}
\usepackage{amssymb}
\usepackage{bm}

\newcommand{\vk}{\mathbf{k}}

\newcommand{\vq}{\mathbf{q}}
\newcommand{\vQ}{\mathbf{Q}}

\renewcommand{\vr}{\mathbf{r}}
\newcommand{\vR}{\mathbf{R}}

\newcommand{\vM}{\mathbf{M}}

\newcommand{\ds}{\downarrow}
\newcommand{\be}{\begin{eqnarray}}
\newcommand{\ee}{\end{eqnarray}}
\newcommand{\p}{\partial}

\newcommand{\vL}{\mathbf{L}}
\newcommand{\vj}{\mathbf{j}}

\def\ket#1{|#1\rangle}

\def\bra#1{\langle #1 |}
\def\ep#1{\langle #1 \rangle}

\begin{document}
\title{Theory of Polarized Neutron Scattering in the Loop Ordered Phase of Cuprates}

\author{Yan He and C.M. Varma}

\affiliation{Department of Physics, University of California, Riverside, CA}
\date{\today}
\begin{abstract}

The collective modes observed in the loop-current ordered state in under-doped cuprates by polarized neutron scattering require that the ground state is a linear combination in each unit-cell  of the four basis states which are the possible classical magnetic moment configurations in each unit-cell. The direction of such moments is in the c-axis of the crystals. The basis states are connected by both time-reversal as well as spatial rotations about the center of the unit-cells. Several new features arise in the theory of polarized neutron scattering cross-section in this situation which appear not to have been encountered before.  An important consequence of these is that a finite component transverse to the classical magnetic moment directions is detected in the experiments. We show that this transverse component is of purely quantum-mechanical origin and that its direction in the plane normal to the c-axis is not detectable, even in principle, in experiments, at least in the quantum-mechanical model we have adopted. We estimate the direction of the ``tilt" in the moment, i.e. the ratio of the transverse component to the c-axis component, using parameters of the ground state obtained by fitting to the observed dispersion of the collective modes in the ordered state. We can obtain reasonable agreement with experiments but only by introducing a parameter for which only an approximate magnitude can be estimated.  Approximate calculations of the form-factors are also provided.
\end{abstract}

\maketitle

\section{introduction}

\begin{figure*}
\centerline{\includegraphics[width=0.9\textwidth]{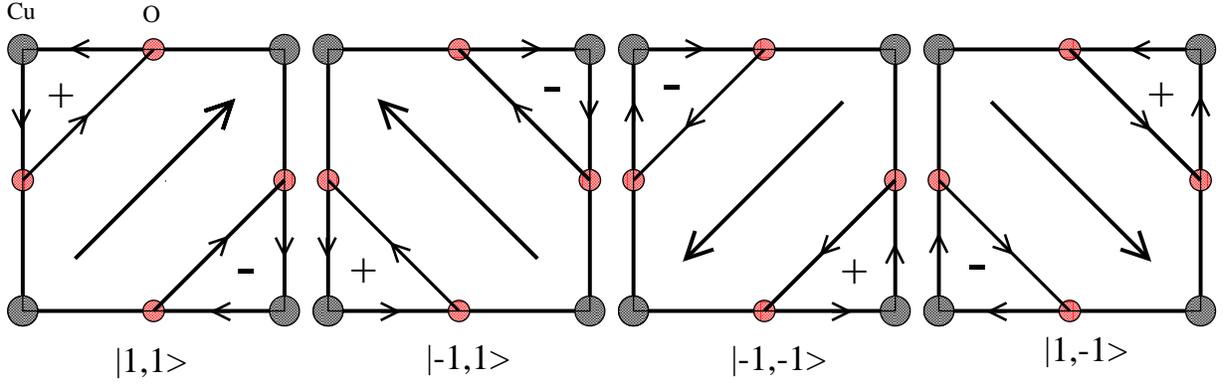}}
\caption{The four Possible ``classical" domains of the loop ordered state are shown. In the classical ordered phase, one of these configurations is found in every unit-cell.}
\label{domains}
\end{figure*}

Polarized elastic neutron scattering experiments \cite{loop-order-expt, bourges-sidis-colloq} and dichroic ARPES \cite{kaminski} have revealed that the pseudo-gap phase of the cuprates have a long-range magnetic order which breaks time-reversal symmetry without breaking translational symmetry of the lattice. It is a ${\bf Q}=0$ staggered order with zero net-moment in each unit-cell. Its geometric arrangement is consistent with the order of a pair of oppositely directed fluxes due to current loops formed in the o-cu-o links in each unit-cell \cite{cmv-2006}. Classically such an order has four possible domains as shown in Fig. \ref{domains}. These domains are specified by the directions $(\pm 1, \pm 1)$, that the order parameter ${\bf \Omega}$ makes with respect to the x and y-axes of the crystal. The order parameter is an {\it anapole} \cite{zeldovich, shekhter} given by
\be
\label{anapole}
{\bf \Omega} = \int_{cell} d{\bf r} \big({\bf L}({\bf r}) \times {\bf r}\big),
\ee
where ${\bf L}({\bf r})$ is the magnetic moment in the unit-cell at the point ${\bf r}$. Such an order, for any of the four possible domains,  has orbital magnetic moments ${\bf L}$ pointed in directions along or opposite the c-axis of the crystals. However, polarization analysis of the neutron scattering \cite{loop-order-expt} has shown that this is not true. The direction of the moments, interpreted according to the classic theory of polarized neutron scattering \cite{lovesey, squires}, makes a large angle with respect to the c-axis \cite{li-greven-prb}; the direction along the plane is not revealed due to  the multi-domain nature of the crystals and/or the multi-domain nature of the order {\it or} as we will show here due to its quantum-mechanical nature.

Polarized inelastic scattering has also discovered \cite{coll-modes}, \cite{mook} two branches of weakly dispersive collective modes in the same temperature region as the magnetic order and with an intensity as a function of temperature compatible with it. Such collective modes can only be understood as quantum-fluctuations of the observed order, just as is true for the one branch of collective modes in the transverse field Ising model \cite{deGennes}. In the paper preceding this \cite{he-cmv-prl}, hitherto referred to as I, we have introduced a quantum-mechanical model for the observed order and calculated  the simplest quantum-mechanical ground state of the model as well as the collective modes. The ground state is a product over the unit-cells of a sum over the four classical configurations in each unit-cell, depicted in Fig. \ref{domains} and given with {\it a particular choice of phases} by
\be
\ket{G}&=& \prod_i\Big(\cos^2\frac{\theta}2\ket{1,1}_i
+\cos\frac{\theta}2\sin\frac{\theta}2
(\ket{1,-1}_i\nonumber\\
& &\qquad+\ket{-1,1}_i)+\sin^2\frac{\theta}2\ket{-1,-1}_i\Big)
\label{GS}
\ee
$\theta$ is a parameter which has been determined by fitting the calculated collective mode dispersions to the experiments. $\ket{\pm 1, \pm 1}_i$ refer to the four configurations in a unit-cell $i$.
As discussed in detail in I, and will be summarized below, one can make unitary transformations consistent with the symmetry of the problem, which introduce other operators in the ground state, and give a corresponding ground state wave-function, which in general is a linear combination of the basis states with complex coefficients. We show in this paper that polarized neutron scattering from such a ground state requires a quantum-mechanical description of scattering of neutron of the quantum magnetic moments, whereas the traditional method considers the problem as a quantum-mechanical scattering of neutron from a classical magnetic field due to the ordered magnetic moments (and in some cases their zero-point fluctuations which only cause Debye-Waller like corrections in amplitudes). We show that in this situation, the observed ``tilting" of the moments is a purely quantum phenomena. One can only deduce by neutron scattering, or by any other experiment, the component of the magnetic moment along the c-axis of the crystal and that perpendicular to it but not the two orthogonal components of the latter. We compare various aspects of the experiment with our calculations. Aspects of the symmetry of the observations all appear to be well reproduced. The quantitative magnitude of the ``tilt"  and its variation with the Bragg-vector can only be reproduced by introducing a free parameter, which can only be estimated approximately.

The basis states in (\ref{GS}) may be taken to be the eigenstates of the orbital magnetic moment operator ${\bf L}_3$, with the $3$-axis identified with the c-axis of the crystal. An important aspect of the problem is to specify the kinetic energy term which mixes the four basis states in a cell $i$ to give (\ref{GS}).  This is important because there is obviously no orbital moment vectors pointing in the plane at the four locations indicated in the figure
(\ref{domains}) because current flow only in the planes is assumed. We will show that an operator with the right commutation rules for an angular momentum exists in the problem but its physical basis is a fluctuating current loop between the four oxygen atoms around a cu atom in each unit-cell. Such an operator occurs naturally in the microscopic theory of loop currents in the cuprates \cite{aji-shekhter-varma-pr} and leads among other things to the marginal fermi-liquid \cite{kotliar, cmv-prl} in the normal state in the quantum-critical regime.

This paper is organized as follows. In the next section, we summarize the quantum Ashkin-Teller model introduced in I and cast it in the basis of local angular momentum operators which are more useful to discuss neutron scattering. We will also review the transformation properties of these operators to show that, given the quantum Ashkin-Teller model, the direction of the ``tilt" in the x-y plane cannot be determined. In the following section, we discuss neutron scattering and its polarization dependence and show that besides the usual matrix element for polarized neutron scattering, there exists also another matrix element due to the finite extent of the current loops. In calculating the neutron scattering intensity, we first consider the moments as point objects at the four sites and subsequently improve the calculation by considering the finite extent of the current loops. This also allows us to estimate the form factors or the momentum dependence of the scattering at the Bragg vectors.

In our conclusions, we discuss also alternate ways of obtaining the "tilts" and show that they are not consistent with the qualitative features of the experiments.



\section{Loop-Current Magnetic Order}

In I we have fully described the symmetries of the quantum Ashkin-Teller model with which the collective modes of the ordered loop current states are described and compared with experiments. To calculate neutron scattering, we will proceed in two steps. First, we will stay with the abstract representation given in I. To reproduce only the correct symmetry of the magnetic order, the orbital moments have been represented as point objects located in the centroid of the triangular loops of Fig. (\ref{domains}). The locations are labeled by $\vR_{i,a}=\vR_i^0+\vR_a$, $a = 1,..4$. Here $i$ denote the lattice sites, $\vR_i^0$ is the position of the center of the unit cell and $\vR_a$ is the relative position of each local moments in the unit cell. We take $\vR_1=(r_0,r_0)$, $\vR_2=(-r_0,r_0)$, $\vR_3=(-r_0,-r_0)$ and $\vR_4=(r_0,-r_0)$ with a value $r_0$ smaller than 1/2 the lattice constant.
The 4 loop current states are labelled by the eigenvalues of $\sigma_z$ and $\tau_z$ of the classical Ashkin-Teller (AT) model. At the classical level, for state $\ket{1,1}$ there is a magnetic moment perpendicular to the copper oxygen plane pointing up located at $\vR_1$ and another pointing down at $\vR_3$. Also there are  zero moments  at $\vR_2$ and $\vR_4$. All the other 3 loop current states are can be obtained by sequentially rotating $\ket{1,1}$ by $\pi/2$ and will be denoted by $\ket{-1,1}$, $\ket{-1,-1}$ and $\ket{1,-1}$. This labeling is also consistent with the direction of the anapole vector ${\bf \Omega}$. The four states $(\pm 1, \pm1)$ are of-course globally orthogonal. We assume that this is also true of the four local states in any unit-cell $(\pm 1, \pm1)_i$, as also each of them between different cells. This is no different than, say, what is done with respect to the local moments formed from the collective degrees of freedom of fermions in itinerant anti-ferromagnets or ferromagnets for regions of frequency and momenta where their exceptions do not overlap much the incoherent fermion excitations.

This is  adequate to get the symmetries of the neutron scattering intensity for different momentum transfer ${\bf Q}$ and different measured initial and final neutron polarizations. But it is not adequate to give the form-factor and relative intensities at different ${\bf Q}$ and polarizations. Calculating such information is a very formidable task. In a second step, we will however attempt this in an approximate way by introduce the wave-functions responsible for generating the orbital moments as well as the representation of the kinetic energy in terms of current-operators.  Besides quantitative verisimilitude, this affords some insight into the interesting new physics in the kinetic energy terms.

The Hamiltonian derived in I to calculate the collective modes was written in the space basis of the states $(\pm 1, \pm1)$,  and equivalently in the direct product basis with 4-dimensional vectors. The transformation from one to the other is,
\be
\ket{1,1}_i \equiv \ket{1 0 0 0}_i;~ \ket{1,-1} \equiv \ket{0 1 0 0}_i, etc.
\ee
The classical Ashkin-Teller model in this basis is given by
\be
H_{AT}=-\sum_{\langle i,j\rangle}[J_1 S^3_i S^3_j+J_2T^3_i T^3_j+J_4 K^{33}_i K^{33}_j]
\ee
 The quantum terms causing a transition between the four states which are rotated with respect to each other in the direction of ${\Omega}$ by $\pm \pi/2$. are given {\it in the same choice of gauge as (\ref{GS})} by
\be
H_Q = \sum_i t(S^1 + T^1) + t' K^{11}
\ee
where $S^i, T^i , K^{ii}$ are matrices in $SU(4)$ space, specified in I, where their  commutation rules are also given. The  Hamiltonian  $H_{AT} + H_Q$ was used to derive the collective modes in I and to get the ground-state wave function $\ref{GS}$.
The terms $K^{11}_i$ which causes rotation by $\pm \pi$,  are unimportant for our purposes here because they cause change in angular momentum by 2 and therefore do not couple to neutrons, which can change angular momentum only by 1 in the weak scattering limit.

For calculating neutron scattering, it is more convenient to define a basis set given in terms of the orbital moment operators at the 4 location in a cell.
Since the local moments are generated by orbital loop current they should be considered in the representation for spin 1. The three components of the effective moment will be denoted by ${\bf L} = (L^1, L^2, L^3)$. Normally, one would represent ${\bf L}$ by the spin-1  representation of $SU(2)$, i.e. by a three dimensional representation, with eigenvalues say of $ \pm 1, 0$. But we have four states per unit-cell in the loop current model. As explained in the Appendix \ref{4d}, the three dimensional representation is inadequate to the present case  and a four dimensional representation of the spin 1 states must be used. This representation is given by

\be
\label{L=1}
&&L^1=\frac{1}{2}\left(\begin{array}{cccc}
0&1&1&0\\
1&0&0&1\\
1&0&0&1\\
0&1&1&0
\end{array}\right),\quad
L^2=\frac{1}{2}\left(\begin{array}{cccc}
0&-i&-i&0\\
i&0&0&-i\\
i&0&0&-i\\
0&i&i&0
\end{array}\right),\nonumber\\
&&L^3=\left(\begin{array}{cccc}
1&0&0&0\\
0&0&0&0\\
0&0&0&0\\
0&0&0&-1
\end{array}\right)
\ee
This representation is equivalent to a representation in the basis of  2 independent set of spin-1/2 operators. This is quite natural to use for the quantum Ashkin-Teller model.  We will verify that they satisfy the $SU(2)$ algebra $[L^i,L^j]=i\epsilon^{ijk}L^k$. This is a reducible representation with two (orthogonal) $L_3=0$ states we denote as $\ket{0^1}$ and $\ket{0^2}$. The loop current basis states in each unit-cell in this representation have the following bases in each unit-cell:
\be
\label{basis1}
&&\ket{1,1}=(\ket{1}_{\vR_1},\ket{0^1}_{\vR_2},\ket{-1}_{\vR_3},-\ket{0^2}_{\vR_4})\\
&&\ket{-1,1}=(\ket{0^1}_{\vR_1},\ket{-1}_{\vR_2},-\ket{0^2}_{\vR_3},\ket{1}_{\vR_4})\\
&&\ket{-1,-1}=(\ket{-1}_{\vR_1},\ket{0^2}_{\vR_2},
\ket{1}_{\vR_3},-\ket{0^1}_{\vR_4})\\
&&\ket{1,-1}=(\ket{0^2}_{\vR_1},\ket{1}_{\vR_2},-\ket{0^1}_{\vR_3},\ket{-1}_{\vR_4})
\ee

Here $\ket{i}_{\vR_j}$ stands for the eigenvector of $L_3$ with eigenvalue $i = \pm 1$ or $0$   at location $j = 1, ...4$  in a unit-cell. The phase factors $\pm 1$ in front of the states $\ket{0}_{\vR_j}$ are picked so that the states in the left of Eq. (\ref{basis1}) have zero net moment in a cell.

While, since it has a classical analog, it is perfectly clear what $L^3_{i,s}$ physically means, the same cannot be said of
$L^1_{i,s}$ and $L^2_{i,s}$.
Obviously, these are not proportional to angular momentum operators at the sites $(i,s)$, since the currents are required to flow only in the plane. We shall show that there exist operators which have off-diagonal matrix elements in the basis $(\pm 1, \pm1)$ , so that using (\ref{basis1}), we shall find (in general complex) matrix elements between the eigenstates with eignevalues $\pm1$ and $0$
of the operator $L^3_{i,s}$. We shall define $L^1_{i,s}$ and $L^2_{i,s}$ through such matrix elements.




\subsection{Allowed Unitary transformations on the Quantum Ashkin-Teller Model}

The unitary  transformations described in I are equivalent to rotations in $L$-space, as we will explicitly show below. Apart from the crystalline $z$-axis which defines the direction of the moments of the classical problem, there is nothing in the Hamiltonian derived in I for the problem which involves the crystalline $x$- and $y$-axis (except that they be orthogonal to the $z$-axis).  Therefore the direction $3$ of $L^3$ may be identified as the $z$-axis leaving the choice of the 1 and 2 axis with respect to the $x$ and $y$ crystalline-axes undetermined.  This situation is similar to the traverse field Ising model in which the transverse field is self-generated. One can then choose it any direction normal to the Ising axis with identical experimental results. (Only if the transverse field is an externally applied field can one find in experiments that the direction of the moments is tilted from the Ising axis towards the external field axis.)

We have shown in I that the classical Ashkin-Teller model has a continuous symmetry $U(1)_{S^3}\times U(1)_{T^3}\times U(1)_{K^{33}}$ which is the rotation around operators $S^3$, $T^3$ and $K^{33}$. The rotation matrix is given by
\be
\label{gauge-tr.}
U(\theta)=e^{i\theta_1S^3/2}e^{i\theta_2T^3/2}e^{i\theta_3K^{33}/2}
\ee
The classical AT model is invariant under this transformation. It has been shown that the quantum terms introduced depend on the transformation $U$. At the same time, this transformation  changes the wave functions. For a general wave function $\ket{\psi}=a\ket{1,1}+b\ket{1,-1}+c\ket{-1,1}+d\ket{-1,-1}$ with four complex coefficients $a,b,c,d$, we have
\be
\ket{U\psi}&=&ae^{i\phi_1}\ket{1,1}+be^{i\phi_2}\ket{1,-1}\nonumber\\
& &+ce^{i\phi_3}\ket{-1,1}+de^{i\phi_4}\ket{-1,-1}
\ee
with $\phi_1=\frac{\theta_1+\theta_2+\theta_3}{2}$, $\phi_2=\frac{\theta_1-\theta_2-\theta_3}{2}$,
$\phi_3=\frac{-\theta_1+\theta_2-\theta_3}{2}$ and
$\phi_4=\frac{-\theta_1-\theta_2+\theta_3}{2}$. Thus $U$
puts different phase factors on each of the 4 classical states with the sum of all phase factors restricted to zero.

One should note the special case $\theta_1=\theta_2=0$, when this unitary transformation is simply a rotation of ${\bf S}$ and equivalently of ${\bf L}$ in the x-y plane:
\be
\label{res.gauge}
UL^1U^{\dagger}=\cos\theta_3L^1-\sin\theta_3L^2, \\
UL^2U^{\dagger}=\sin\theta_3L^1+\cos\theta_3L^2.
\ee
So, given the quantum Ashkin-Teller model, the direction of the moments perpendicular to the z-axis is un-determinable.

We now show more generally that ${\bf L}_i$ have the properties of angular momentum operators.
As already discussed in \cite{he-cmv-prl}, by making use of Eq.(\ref{basis1}) etc, one can express the local spin operator $\vL$ in terms of the Ashkin-Teller model operators of $S^i$, $T^i$ (defined in \cite{he-cmv-prl}) as follows
\be
&&L^x_{\vR_1}=L^x_{\vR_2}=\frac{1}{2}(S^1+T^1),\nonumber\\
&&L^x_{\vR_3}=L^x_{\vR_4}=-\frac{1}{2}(S^1+T^1),\nonumber\\
&&L^y_{\vR_1}=L^y_{\vR3}=\frac{1}{2}(S^2+T^2),\nonumber\\
&&L^y_{\vR_2}=L^y_{\vR4}=\frac{1}{2}(S^2-T^2),\nonumber\\
&&L^z_{\vR_1}=-L^z_{\vR3}=\frac{1}{2}(S^3+T^3),\nonumber\\
&&L^z_{\vR_2}=-L^z_{\vR4}=\frac{1}{2}(S^3-T^3)\nonumber
\ee
Using the commutation relations given for $S^i, T^i$ in I, it is easy to verify that the $\vL$ satisfy the $SU(2)$ algebra (up to an overall minus sign). Therefore, $\vL$ can be regard as an angular momentum operators.

\section{Polarized Neutron Scattering}

The neutron scattering Hamiltonian is
\be
\label{neutron}
H_{int} = \int d{\bf r} {\bf B}({\bf r}) \cdot {\bf \sigma}({\bf r}).
\ee
$\sigma({\bf r})$ is the spin of the neutron at point $({\bf r})$ in the crystal where the magnetic field {\it operator} is ${\bf B}({\bf r})$. The source of the magnetic field are the magnetic moment {\it operators} due to spin or in our case orbital moments ${\bf L}(\vR_i^0 + \vR_{a})$ at locations $(\vR_i^0 + \vR_{a})$. One can Fourier transform (\ref{neutron}) and rewrite it in terms of the magnetic moments ${\bf L}$ at the momentum transfer ${\bf Q}$ as \cite{lovesey, squires}
\be
\label{neutron-Q}
\sum_ae^{i\vQ\cdot\vR_a}F_a(\vQ)\vL_{a,\bot}(\vQ)\cdot\sigma(\vQ)
\ee
Here ${\bf L}_{a,\bot} ({\bf Q}) = {\bf L}_a - ({\bf L}_a\cdot {\hat{\bf Q}}){\hat{\bf Q}}$ is the component of ${\bf L}_a$ perpendicular to ${\hat{\bf Q}}$ and $F_a(\vQ)$ is the form factor.

It is important to discuss how the directions of the Pauli-matrices $\sigma$ are fixed in the usual situation in which the directions of ${\bf L}$ are known with respect to the crystalline axes and the difference in the present case. The quantization axis of the neutron spin is fixed externally to the sample by applying a (small) magnetic field in a specific direction, with respect, say to the momentum transfer direction ${\hat{\bf Q}}$ of the neutron. This fixes $\sigma_3$ with respect to the crystalline axes and the experiment is done with various choices of $3$ with respect to
${\hat{\bf Q}}$. The other directions $1$ and $2$ are then fixed through knowing the direction of ${\bf L}$ with respect to the crystalline directions and the use of the dot-product in (\ref{neutron}). In effect, ${\bf B}$ can be treated classically in such situations.

This is to be contrasted with the present situation in which the basis vectors of the ground state $(\pm 1, \pm 1)$ specify only the direction of the orbital moment, up or down (or zero) as being along the normal to the cu-o planes denoted here by the $z-axis$. Taking matrix elements of ${\bf B}$ in the ground state ({\ref{GS}) leads to off-diagonal terms in these basis vectors. As noted this cannot be specified as a magnetic field operator generated by magnetic moments in specific directions with respect to the crystalline axes; all that can be said is that the off-diagonal matrix elements are matrix elements of a magnetization operator orthogonal to the direction ${\hat z}$ with which the basis vector are specified. A purely real ground state wave-function means that only matrix elements of $S^1$ and $T^1$ generate the off-diagonal elements and so only  ${\bf L}^1_{\bot} ({\bf Q})$ enters in in (\ref{neutron-Q}). Correspondingly, only
$\sigma_1$ appears in (\ref{neutron-Q}). There is no way to fix $1$ with respect to the crystalline axes. If however one used a more general choice of the wave-function so that it is complex,
${\bf L}_{\bot} ({\bf Q})$ are determined by matrix elements of $S^1,T^1$ as well as $S^2, T^2$, and correspondingly $\sigma_2$ enters in the calculation. The final answer for the spin-flip cross-sections of the neutrons cannot (and does not) depend on the choice of the wave-function, nor can the directions $1$ and $2$ be determined with respect to the crystalline axes.

The difference in our case from the traditional case arises from the fact that ${\bf L}$ does not come physically from an atomic orbital moment where the three different components of the orbital angular momentum can be defined with respect to the crystalline axes. Rather, in our case only the z-component is defined in the basis; the mixing in the ground state of the basis is due to a transverse field operator as discussed above. The physical basis for the transverse field will be specified below.

Our purpose is to interpret experiments which deduce everything from measuring
(functions of) the matrix elements of $\sigma({\bf Q})$ through three different choices of the quantization axis $3$ with respect to the ${\bf Q}$. From (\ref{neutron-Q}), it follows that when the polarization of the neutron, i.e the direction 3 of $\sigma$ is chosen parallel to ${\bf Q}$, there is only spin-flip scattering while for any other choice there is both spin-flip and spin-nonflip scattering. We have identified $L^3$ as proportional to $L_z$. As explained, we have a freedom of choice of  rotating the 1 and 2 directions of the neutron $\sigma$ by any arbitrary angle about its chosen 3-axis.  It thus follows that for any choice of ${\bf Q}$ and the neutron polarization, one can never determine ${\bf L}_x$ and ${\bf L}_y$. One can only determine ${\bf L}_z$ and the component of ${\bf L}$ perpendicular to it, which we will call ${\bf L}_t$. Having shown this, we can do the calculation in the simplest choice in which the wave-function is real and only the operators $S^1, T^1$ and $K^{33}$ appear in the Hamiltonian.

The situation may be contrasted with the case when the direction of order of the system is fixed by, for example, crystalline anisotropy as in the anisotropic Heisenberg model. The order parameter $\langle{\bf M}\rangle$ is then fixed with  respect to the crystalline axes and may be regarded as a classical source for a classical ${\bf B}({\bf r})$ in Eq. (\ref{n-ham}). The magnitudes of two of the components of $\langle{\bf M}_x, {\bf M}_y, {\bf M}_z \rangle $ can then be determined by measuring the neutron scattering cross-sections by polarizing the neutron beam in two different directions with respect to a momentum transfer ${\bf Q}$ and using Eq. (\ref{neutron-Q}).  One can then change ${\bf Q}$ and repeat the measurement to determine all the three directions of $\langle{\bf M}\rangle$ for simple magnetic order (or measure at other ${\bf Q}$ for more complex order.)
If  one has an Ising model in a external transverse field, the direction of ${\bf M}$ is similarly fixed.

The problem discussed above is different also from problems of scattering neutrons in the quantum Heisenberg antiferromagnets, where the kinetic energy terms in the Hamiltonian are quadratic operators in the spins, as opposed to the present case where they are linear. In such cases, quantum-mechanics only induces a reduction in the ordered spin-moment without changing its orientation through a Debye-Waller factor due to zero point spin fluctuations while transferring weight to an incoherent background.


\section{Matrix Elements for Neutron Scattering}

We shall show here that there are two kinds of matrix elements  in scattering of neutrons due to the linear combination in the ground state (\ref{GS}). This can be seen most clearly from the rotational and time-reversal properties of the basis states written in the form of Eqs. (\ref{L=1}). The basis states are connected though what we might call the time-reversal part of the dipole Hamiltonian which involve $L^+\sigma^- + L^-\sigma^+$. This is the usual scattering. But as already mentioned the basis states also go to each other under successive $\pi/2$ and $\pi$-rotations in real space through the axis normal to the plane at the center of a cell. We show here that the dipole interaction (\ref{n-ham}) has a finite projection to such rotation operators also. We will call these matrix elements of the "rotational" kind.  The matrix elements from a given initial basis state and a given final basis state for both kinds must be summed and then squared to get the scattering cross-section.

Since the local loop current is actually an extended object, the neighboring loop currents although orthogonal have a finite matrix element through the spatial dependence of the dipole interactions Hamiltonian:
\be
H_{int}(\vR)&=&\frac{\bm{\sigma}\cdot\vL
-3(\bm{\sigma}\cdot\widehat{(\vr-\vR)}
(\vL\cdot\widehat{(\vr-\vR)}}{|\vr-\vR|^3}\nonumber\\
&=&-\bm{\sigma}\cdot\nabla_{\vr}\times
[\vL\times\nabla_{\vr}\frac{1}{|\vr-\vR|}]
\label{n-ham}
\ee
Here $\widehat{(\vr-\vR)}$ is the unit vector along $(\vr-\vR)$. $\vr$ and $\vR$ is the position vector of neutrons and local moments.

Let $\ket{\psi_{a}}$ denote one of the four loop current state $\ket{\pm1,\pm1}$ and $\ket{\psi_a,\vR_i}$ denote the local moment state of $\psi_{a}$ located at $\vR_i$. This local moment state can be written as the direct product of the coordinate part and a "moment" part as
$\ket{\psi_a,\vR_i}=\ket{\phi(\vR-\vR_i)}\ket{\psi_a,\vR_i}_s$. Here the moment part $\ket{\psi_a,\vR_i}_s$ are the four possible states $\ket{\pm1}$ and $\ket{0^{1,2}}$. The coordinate part $\phi(\vR-\vR_i)$ describes the finite size distribution of the magnetic moment centering around $\vR_i$. For orbital moments, $\phi$ may be taken to be the real part of the wave-function while $\psi$ may be taken to be a phase varying around the loop for the loop currents. For  $\phi(\vR-\vR_i)=\delta(\vR-\vR_i)$, there are only the usual matrix elements of the dipole interaction between the basis states of \ref{GS}) because
 \be
&&\int d^3R\,\phi(\vR-\vR_i)H_{int}(\vR)\phi(\vR-\vR_j)\nonumber\\
&&=\int d^3R\,\delta(\vR-\vR_i)H_{int}(\vR)\delta(\vR-\vR_j) \propto \delta_{ij}.
\ee
We will call the matrix elements of the dipole interactions for $(i=j) $ the matrix elements of the "spin" kind.

But for finite size $\phi$'s, the above matrix element is in general not zero. The detailed form of this function $\phi(\vR)$ is not very important to us. Here we also assume that the distribution function satisfies the orthogonal relations as
$\int d^3R\,\phi(\vR-\vR_i)\phi(\vR-\vR_j)=\delta_{ij}$.

Then such matrix element are
\be
\label{rot-me}
& &\bra{\psi_a,\vR_i}H_{int}(\vR)\ket{\psi_b,\vR_j}\nonumber\\
&=&-\int d^3R\,\phi(\vR-\vR_i)\bm{\sigma}\cdot{\bf B}\phi(\vR-\vR_j)\nonumber\\
& &{\bf B}=\nabla_{\vr}\times[\vL_{ij}\times
\nabla_{\vr}\frac{1}{|\vr-\vR|}]\nonumber
\ee
Here $\vL_{ij}=\bra{\psi_a,\vR_i}_s\vL\ket{\psi_b,\vR_j}_s$. We will call them matrix elements of the ``rotational kind".

\subsection{Matrix element of the ``spin" kind}

The ground state expectations of ${\bf L}_i$ using Eq. (\ref{basis1}) is,
\be
\label{exval}
&&\langle{\bf L}_i(\vR_1)\rangle=(\sin\theta,\cos\theta),\\
&&\langle{\bf L}_i(\vR_2)\rangle=(\sin\theta,0)\\
&&\langle{\bf L}_i(\vR_3)\rangle=(-\sin\theta,-\cos\theta),\\
&&\langle{\bf L}_i(\vR_4)\rangle=(-\sin\theta,0),
\ee
where now the three components refer to the ``directions" $\hat{t}$ and $\hat{z}$, respectively.

The calculation in Eq. (\ref{exval}) is only for one unit cell. The magnetization in the lattice just repeats the same result in each unit cell.
$$
\vL(\vQ)=\sum_{\vR_{ia}}\vL(\vR_{ia})e^{-i\vQ\cdot\vR_{ia}}
$$
and the magnetization is given by
\be
&& L_t(\vQ)=\sum_{n}\sin\theta(e^{-i\vQ\cdot\vR_1}+e^{-i\vQ\cdot\vR_2}\nonumber\\
&&\qquad\qquad-e^{-i\vQ\cdot\vR_3}-e^{-i\vQ\cdot\vR_4})
\delta(\vQ-\bm{\tau}_n)\label{Mx}\\
&& L_z(\vQ)=\sum_{n}\cos\theta(e^{-i\vQ\cdot\vR_1}-e^{-i\vQ\cdot\vR_3})
\delta(\vQ-\bm{\tau}_n)\nonumber\\
\label{Mz}
\ee
 Here $\bm{\tau}_n$ is the reciprocal lattice vector of the lattice. Suppose the unit vector of transfer momentum is $\hat{\vq}$, then the scattering intensity is
\be
I(\vQ)\propto|\langle\vL_{\perp}({\bf Q})\rangle|^2
\label{spin-flip}
\ee
Since we treat the orbital current loop as a point-like spin, the scattering amplitude is a constant as function of $\vQ$. So, this calculation is not designed to give the structure factor correctly.

Suppose this is the only contribution to scattering. Then at any  $\vQ=(a^*,0,0)$, barring multiple cu-o layers per unit-cell (which we will have to consider for $YBa_2Cu_2O_{6+\delta}$, the "tilt" angle would be determined by
$$
\frac{|L_t|}{|L_z|}\approx\frac{2\sin\theta\cdot 2\sin(Qr_0)}
{\cos\theta\cdot 2\sin(Qr_0)}
$$
For the fitted parameters from calculation of the collective modes in I, we have $\sin\theta=-0.26$, so the tilt angle is only about $28^{\circ}$.

\subsection{Matrix elements of ``Rotational" kind}

We now consider the second kind of scattering starting from Eq. (\ref{rot-me}).

We can make use of the fact $\frac1r=\int d^3q e^{i\vq\cdot\vr}\frac1{q^2}$ to transform the above result to momentum space.
\be
&&\bra{\psi_a,\vR_i}H_{int}(\vR_1)\ket{\psi_b,\vR_j}\nonumber\\
&&=\bm{\sigma}\cdot\int d^3Q e^{i\vQ\cdot\vr}\,\hat{\vQ}\times(\vL_{ij}\times\hat{\vQ})S^{ij}(\vQ)\\
&&\mbox{with}\qquad S_{ij}(\vQ)=\int d^3R\,\phi(\vR-\vR_i)\phi(\vR-\vR_j)
e^{-i\vQ\cdot\vR}\nonumber
\ee
Now we can expand the exponential factor and the leading nonzero term is
\be
S_{ij}(\vq)\approx-i\int d^3R\,\phi(\vR-\vR_i)\phi(\vR-\vR_j)(\vQ\cdot\vR)
\ee
To be specific, we consider $S_{12}(\vq)$ first. For simplicity, we assume that the distribution is isotropic $\phi(\vR)=\phi(|\vR|)$. Since $\vR_1=(r_0,\,r_0)$ and $\vR_2=(-r_0,\,r_0)$, then it is easy to see that $\phi(|\vR-\vR_1|)\phi(|\vR-\vR_2|)$ is even function about $R_x$ and $R_z$. Thus we have
\be
S_{12}(\vQ)&\approx&-iq_y\int d^3R\,\phi(\vR-\vR_1)\phi(\vR-\vR_2)R_y\nonumber\\
&=&-iQ_yC\\
S_{23}(\vQ)&\approx&-iq_x\int d^3R\,\phi(\vR-\vR_2)\phi(\vR-\vR_3)R_x\nonumber\\
&=&iQ_xC\\
S_{34}(\vQ)&\approx&-iq_x\int d^3R\,\phi(\vR-\vR_3)\phi(\vR-\vR_4)R_y\nonumber\\
&=&iQ_yC\\
S_{41}(\vQ)&\approx&-iq_x\int d^3R\,\phi(\vR-\vR_4)\phi(\vR-\vR_1)R_x\nonumber\\
&=&-iQ_xC
\ee
where $C=\int d^3R\,\phi(|\vR-\vR_1|)\phi(|\vR-\vR_2|)R_y\nonumber$.
We see that $C$ is of $O(r_0)$. We will find the coefficient only by fitting to the data.

Note that this new term is also proportional to ${\bf L}_{\bot}({\bf Q})$, just as the traditional matrix elements for polarized neutron scattering. But the momentum dependence is different from the classical neutron diffraction expression. We can repeat this calculation at all possible neighboring local moments and find similar expressions. Recall that the 4 states are collections of 4 local spin 1 states. The ground state is superposition of the 4 states, thus it is also a collections of 4 local spin 1 states. To compute the total neutron scattering amplitude, we need to evaluate the matrix elements of $\vL$ between all the neighboring local spin 1 states of the ground state. The 4 local spin 1 states of the ground state is given by (Here we only write the spin part)
\be
\ket{G,\vR_1}_s&=&c^2\ket{1}+sc\ket{0^1}+sc\ket{0^2}+s^2\ket{-1}\nonumber\\
\ket{G,\vR_2}_s&=&sc\ket{1}+c^2\ket{0^1}+s^2\ket{0^2}+sc\ket{-1}\nonumber\\
\ket{G,\vR_3}_s&=&s^2\ket{1}-sc\ket{0^1}-sc\ket{0^2}+c^2\ket{-1}\nonumber\\
\ket{G,\vR_4}_s&=&sc\ket{1}-s^2\ket{0^1}-c^2\ket{0^2}+sc\ket{-1}\nonumber
\ee
Here $c = \cos \theta; s = \sin \theta.$ Then it is straightforward to find the following matrix elements for $\vL$
\be
&&\bra{G,\vR_1}_s\vL\ket{G,\vR_2}_s=\Big(\frac{(1+\sin^2\theta)}4,
-\frac i{2}\cos\theta,\frac{\sin(2\theta)}4\Big)\nonumber\\
&&\bra{G,\vR_2}_s\vL\ket{G,\vR_3}_s=\Big(\frac1{2}\cos^2\theta,
-\frac i{2}\cos\theta,-\frac14\sin(2\theta)\Big)\nonumber\\
&&\bra{G,\vR_3}_s\vL\ket{G,\vR_4}_s=-\Big(\frac{(1+\sin^2\theta)}4,
\frac i{2}\cos\theta,\frac{\sin(2\theta)}4\Big)\nonumber\\
&&\bra{G,\vR_4}_s\vL\ket{G,\vR_1}_s=\Big(-\frac1{2}\cos^2\theta,
-\frac i{2}\cos\theta,\frac14\sin(2\theta)\Big)\nonumber
\ee
Other matrix elements are the complex conjugate of the above equations. If we put all the terms together, one can see that the imaginary part cancel out and only the real part contributes. We find the scattering amplitude in the momentum space as
\be
\bra{G}H_{int}(\vR)\ket{G}=2\sum_{(ij)}\bm{\sigma}
\cdot\hat{\vQ}\times(\vL_{ij}\times\hat{\vQ})
S_{ij}(\vQ)
\ee
with $(a,b)=(1,2),(2,3),(3,4),(4,1)$ and $\vL_{ij}=\mbox{Re}\bra{G,\vR_i}_s\vL\ket{G,\vR_j}_s$. Putting all the above results together, we find
\be
&&\bra{G}H_{int}(\vR)\ket{G}=-4iQ_yC\bm{\sigma}\cdot\hat{\vQ}
\times(\vL_{12}\times\hat{\vQ})\nonumber\\
&&\qquad-4iQ_xC\bm{\sigma}\cdot\hat{\vQ}\times(\vL_{41}\times\hat{\vQ})
\ee

Again we can consider the transfer momentum  $\vQ=(a^*,\,0,0)$. The scattering amplitude is
\be
\bra{G}H_{int}(\vR)\ket{G}=-4iQ_xC\bm{\sigma}\cdot\hat{\vQ}
\times(\vL_{41}\times\hat{\vQ})
\ee
which can be rewritten as
\be
\label{mat-ele2}
&&\bra{G}H_{int}(\vR)\ket{G}=\bm{\sigma}\cdot\hat{\vQ}
\times[\vL_{\mathrm{eff}}\times\hat{\vQ}]\\
&& \vL_{\mathrm{eff}}=4iQ_xC\Big(\frac1{2}\cos^2\theta, -\frac14\sin(2\theta)\Big)
\ee

\subsection{Total Matrix Element}
The form, Eq.(\ref{mat-ele2}), has the same structure as the matrix element of the "spin-kind"  Eq. (\ref{spin-flip}). We can combine them together to find the total magnetization as
\be
&&\vL_{\mathrm{tot}}=\vL_0+\vL_{\mathrm{eff}}\\
&&\vL_0=2i\sin(Q_xr_0)(2\sin\theta, \cos\theta)\nonumber\\
&&\vL_{\mathrm{eff}}=4iQ_xC\Big(\frac1{2}\cos^2\theta,-\frac14\sin(2\theta)\Big),\nonumber
\ee
where the two components again refer to $t$ and $z$ respectively.
Here $\vL_0$ is the magnetization form the "spin-type" contribution.

The parameter $r_0$ specifies the location of the moments. For the ground state determined from the collective modes, we have $\sin\theta=-0.26$.
If we assume that  $Q_xC\approx-0.5\sin(Q_xr_0)$, then we find that for transfer momentum $\vQ=(a^*,0,0)$ the tilted angle is about $50^{\circ}$. The experimental results are $55 \pm 7^{\circ}$ and $35 \pm 7^{\circ}$, respectively \cite{loop-order-expt} for $YBaCuO_{6.6}$. But most of the data available  for the collective modes from which the angle $\theta$ is deduced is in $Hg1201$ with a $T_c \approx 61 K$, which from the general phenomenology has properties close to those in $YBaCuO_{6.6}$. Collective modes have also been found in $YBaCuO_{6.6}$ \cite{mook} at similar frequencies but detailed information about the dispersion of the two branches of  collective modes is unavailable. The tilt in $Hg1201$ at $(1,0,1)$ is deduced to be \cite{li-greven-prb}  $45 \pm 20$ within the estimates provided here. We can claim that with $C=O(r_0)$, we get the correct trend and the magnitudes of the tilt in agreement with experiment within the stated error bars. It should be stated that a simplified model is used to calculate the collective modes and the ground state from which $\theta$ is determined. It could easily be $\pm 10^{\circ}$ from that deduced.

\section{Calculation including Form factors}

\subsection{Real space representation of the Operators $(S^1+T^1+S^2+T^2)$}

To calculate the neutron scattering including form factors, one needs the wave-function which have the orbital moments and the microscopic representation of the operators which lead to the produce the admixed ground state wave-function (\ref{GS}). The approximate space representation of the basis states is relatively straight-forward. One may represent them with complex wave-functions carrying currents around the indicated loops as in Fig. (\ref{domains}). The subtle issue is in the representation of the operators $(S^1_i +T^1_i)$, which give
the linear combination of the classical basis states in the ground state in (\ref{GS}). They were introduced at a formal level in I simply because such terms are allowed and because they are necessary to calculate the observed collective modes in the loop ordered state. But what is the physical origin of such quantum terms?  The physical origin was  already derived in connection with calculating the spectra of the collective fluctuations in the quantum-critical regime of the loop ordered phase \cite{aji-shekhter-varma-pr}. A simpler derivation is given in the Appendix \ref{KE-op}. The operator $S^1_i + T^1_i$ is given by the current operator schematically shown in Fig. (\ref{rot-op}) and its hermitian conjugate which has current flowing in the opposite direction. In the continuum limit, this is simply the operator proportional to $\sum_{cells}\nabla\times \vj(\vr)$, with ${\bf r}$ measured from the center of each unit-cell and with cut-off at the boundary. We recall from I that the operator $(S^1_i +T^1_i)$ acting on any of the four states $(\pm 1, \pm 1)$ in a unit-cell admixes the state rotated by $\pm \pi/2$ to it. This operator corresponds to the collective part of the following fermion operators (See Appendix \ref{KE-op})
\be
\label{KE}
\sum_{\vk}e^{i\vk\cdot\vR_i}2i\frac{s_yc_x -s_xc_y}{E_0^2}
(p_{x,k}^{\dagger}p_{y,k}-p_{y,k}^{\dagger}p_{x,k})
\ee
Here $E_0$ is a normalization of the wave-function and $p_{x,\vk}$ and $p_{y,\vk}$ are the fourier transform of the oxygen p-orbital operators in the x-direction and y-direction, respectively around the Cu-atom in the unit-cell $i$.
Such an operator is created microscopically through expressing the nearest neighbor interactions in each unit-cell in terms of current operators and combining such current operators to form closed loops within a unit-cell. Five different point group symmetries result including that of (\ref{KE}) sketched in Fig.(\ref{rot-op}). That this operator admixes the basis states is shown in detail in Appendix \ref{KE-op}.
\begin{figure}
\centerline{\includegraphics[width=0.4\textwidth]{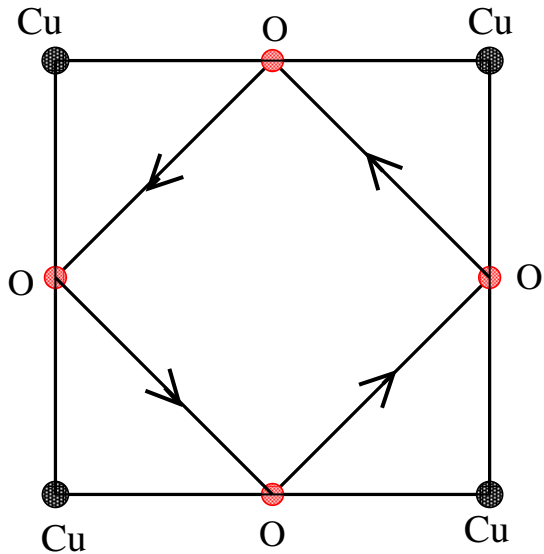}}
\caption{Current pattern of the rotation operator in Eq (\ref{KE}).}
\label{rot-op}
\end{figure}

Given this, we can calculate the form factor of neutron scattering by further approximations to represent the collective states shown in Figures (\ref{domains}) and (\ref{rot-op}).These are described in Appendix \ref{formfac}. We now proceed with the comparison with the experiments in evaluation of the neutron scattering using these results.

\subsection{Results for tilt and form factor}

It easy to see that if we include the form factor effects, $L_z$ decreases faster than $L_1$ as $Q_z$ increases. (See Fig (\ref{Mx-kz})). (Recall that the subscript $1$ in $L_1$ stands for whatever component of ${\bf L}$ perpendicular to the z-axis is being measured.)
We therefore find the tilt angle for $\vQ=(101)$ is larger than $\vQ=(100)$, qualitatively consistent with the experiments \cite{loop-order-expt}. How the tilt angle depends on the momentum is determined in our calculation by the choice of parameters such as $r_s$ and $r_0$,
which is hardly definitive.
\begin{figure}
\centerline{\includegraphics[width=0.5\textwidth]{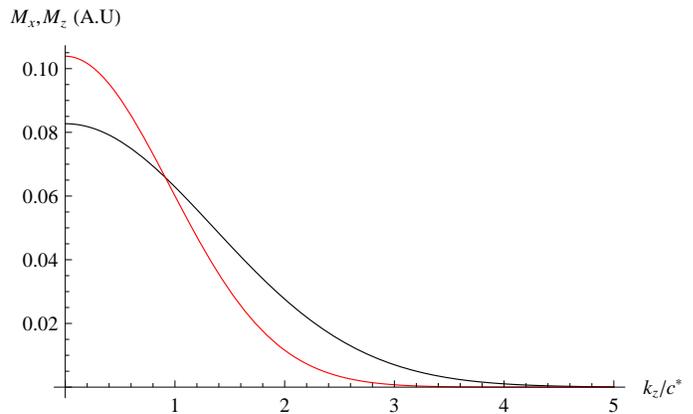}}
\caption{$M_x$ (black curve) and $M_z$ (red curve) as a function of $k_z$.}
\label{Mx-kz}
\end{figure}

For example, as a reasonable guess, we can take $r_0=0.25a$, $r_w=0.175a$. Here $a$ is the lattice constant of xy plane. The tilted angle is quite sensitive to the current width parameter $w$. From experimental data, we know that the intensity of neutron scattering of $L=2$ is half of that for $L=0$. For a Gaussian shape dependence of $Q_z$, we can deduce that $w\approx0.5a$. The half width of the current is comparable to the radius of the current.

For the above choice of parameters, the intersection of these two circles are $\vR_{1a}=(0.238a,0.075a)$ and $\vR_{1b}=(0.075a,0.238a)$. If we use the approximate form factor of Eq. (\ref{Fxak}), then for transfer momentum $\vQ=\frac{2\pi}{a}(0,1,0)$, we find that the tiled angle is $\phi=\arctan(\frac{|L_x|}{|L_z|})\approx64^{\circ}$.

Since the width $w=0.5a$ is not small compared to $a$, we can not use the above approximation. Therefore we have to numerically transform Eq. (\ref{Fxa}) and (\ref{Fza}) to momentum space. For $\vQ=\frac{2\pi}{a}(0,1,0)$, we find that the tiled angle is $\phi\approx 26^{\circ}$. If we also include the ``rotation" kind contribution to the moment from Eq.(\ref{Fxab}) and (\ref{Fzab}) and choose the parameter $C=-0.2a$, we find the total tiled angle is $\phi\approx 38.5^{\circ}$. For YBCO, $c\approx 3a$. Therefore for transfer momentum $\vQ=(0,\frac{2\pi}{a},\frac{2\pi}{c})$, we find the tiled angle $\phi\approx47^{\circ}$.

For YBCO, there are two copper-oxygen planes. Suppose the distance between these two layers are $d$, then we will have an extra structure factor $\cos(Q_z\frac{d}{2})$. This makes the amplitude oscillate with $Q_z$.

\section{Concluding Remarks}

In this paper we have used the result derived earlier in I that the ground state of the loop order state is a linear combination of the classical basis states. This necessitates a new view of calculating the neutron-scattering cross-section, in which one must consider the flip in the neutron spin due also to the matrix elements of flip operators of the quantum-moments in the ground state. The simplest (mean-field) ground state was considered. This does not include the zero-point deviations due to spin-waves. We have calculated these to be a small effect because the collective modes are all at finite energy.

Our method of calculating the neutron scattering cross-section may be useful in other quantum problems, for example the transverse field Ising model, provided the transverse field is not an external field with a specified direction but internally generated. It may also be useful when scattering experiments are done in other loop ordered states such as the anomalous Hall effect and the topological insulator states.

We have been able to show reasonable consistency of the calculation with the experimental results but a parameter $C$ was introduced which we find hard to calculate but can argue only that it should be small compared to 1, as is indeed found. Especially gratifying is that the title angle deduced depends on the Bragg-vector. This would not happen in the traditional usage in which the neutron spin flips quantum-mechanically of a spatially dependent classical magnetic field due to the magnetic order.

It is worthwhile commenting on earlier attempts to understand the ``tilt". One was based on spin-orbit scattering \cite{so-scattering}. This had two difficulties;  there is no such term in $Hg1201$ and the magnitudes do not come out reasonably without giving a scattering at $(2,0,0)$, which is not in agreement with elastic experiments. The other idea is that there is a moment on the triangles made through apical oxygens \cite{weber}. This is allowed by symmetry \cite{shekhter}.  Mean-field calculations do not provide any significant such moment for any reasonable set of parameters. There is a more basic problemm with having any significant amplitude for such a contribution. It is that such moments would provide zero contribution of the moment perpendicular to the z-axis for any scattering vector with ${\bf Q}_z =0$ due to the structure factors in the cu-o lattices with a plane of reflection in the unit-cell perpendicular to the z=axis. A large tilt has been deduced for ${\bf Q} = (1, 0, 0)$.

Our calculations in this paper also have an impact for some other measurements. If
the tilt were due to moments in triangles made through apical oxygens as discussed in the previous paragraph, there should be measurable magnetic fields \cite{kivelson-lederman} at several sites in the lattice, detectable in NMR experiments. With the quantum origin of ``tilt¡± fields, the calculations based on a classical magnetic field tilted in specific directions are not valid. Further consequences of the quantum origin of the ``tilt¡±, for example in NMR experiments
is well worth further investigation.

\acknowledgments
We wish to thank Vivek Aji, Philippe Bourges, Thierry Giamarchi, Martin Greven, Yuan Li, and Yvan Sidis, for discussions of the considerations in this paper and the experimental results and Dr. Stephen Lovesey for a critical reading of the manuscript. We also thank S. Kivelson and S. Lederer for a discussion of the scattering at (1 0 0).

\appendix
\section{Microscopic Theory of the kinetic energy operator $(S^1+T^1+S^2+T^2)$ for the Loop current states}
\label{KE-op}

The loop current states were derived \cite{cmv-2006} starting from the basis of the three-orbital $(d_i, p_{ix}, p_{iy})$ per unit-cell. Besides the kinetic energy operator between the neighboring $d, p_x$ and $p_y$ orbitals, the model includes local repulsion at each site, (the U's) and near neighbor repulsions of charges: $V n_in_j$. The on-site repulsions are assumed to serve only to renormalize the remaining terms in the Hamiltonian in the metallic state. A crucial role is played by the operator for nearest neighbor interactions. For spin-diagonal terms in the operator $n_in_j$, it may be written as
\be
V\sum_{\sigma}n_{i, \sigma}n_{j,\sigma} = V/2(|j_{ij}|^2 - n_i -n_j),
\ee
Here $j_{ij} = i\sum_{\sigma} c_{i,\sigma}^+c_{j,\sigma} + h.c.$ has the operator content of a current. Discarding the one particle terms, a mean-field approximation is made on $|j_{ij}|^2$:
\be
|j_{ij}|^2 = ( r_{ij}^2 + r j_{ij} + r^*j_{ij} + \mbox{fluctuation~ operators}).
\ee
Here $r_{ij}$ is the collective part of $j_{ij}$. Now the effective kinetic energy on the link $(ij)$ is $t_{ij} + i Vr_{ij}/2$ which gives a complex kinetic energy with a phase $\approx r V_{ij}/(2t_{ij})$. Phase differences on links within a unit-cell are combined to form closed loops which have invariant fluxes with different point group symmetries. For non-intersecting loops on the cu-o lattice, there are five and only five such closed loops possible \cite{aji-shekhter-varma-pr} . Two of these which transform as doubly degenerate vectors $E_1$, in the $(x \pm y)$ directions and their time reversed partners form the four flux patterns depicted in Fig.(\ref{domains}). Simple mean-field calculations as well as more elaborate calculations \cite{weber}  show that this is the most stable allowed symmetry and experiments \cite{loop-order-expt} have shown it to be the realized state in all the under-doped cuprates investigated.

The other  current loop which plays a crucial role in our considerations here and in the collective modes of I and played an important role in the calculation of collective fluctuations \cite{aji-shekhter-varma-pr} in the quantum-critical regime is the pattern transforming with the full symmetry of the lattice. It is depicted in Fig. (\ref{rot-op}). We will show that it has local matrix elements between the states $(\pm1, \pm 1)$. Before we show this  it is important to specify some properties of the basis states $(\pm 1, \pm 1)$.

Corresponding to each of the four collective states $(\pm 1, \pm 1)$, there are four kinetic energy Hamiltonians for the fermions:
\be
H(\bf k)=\left(
\begin{array}{ccc}
0 & it_{pd}S_x & it_{pd}S_y\\
-it_{pd}S_x & 0 & t_{pp}s_xs_y\\
-it_{pd}S_y & t_{pp}s_xs_y & 0
\end{array}\right)
\label{H0}
\ee
with $S_x=\sin(k_xa/2+\phi_x)$, $S_x=\sin(k_ya/2+\phi_y)$, $s_x=\sin(k_xa/2)$, $s_y=\sin(k_ya/2)$. Then the four different mean field Hamiltonian correspond to the order parameter
$(\phi_x,\phi_y)=(\pm \Omega,\pm \Omega)$, where $\Omega$ is the magnitude of the order parameter which is determined variationally. The eigenvalues and eigenstates of the three fermion bands in the zeroth order of $t_{pp}/t_{pd}$ are given by
\be
&&E_1=-t_{pd}\sqrt{S_x^2+S_y^2},\qquad E_2=0,\nonumber\\ &&E_3=t_{pd}\sqrt{S_x^2+S_y^2}\nonumber\\
&&\ket{1,k}=\frac{1}{\sqrt{2}}
\Big(-i,\frac{S_x}{S_{xy}},\frac{S_y}{S_{xy}}\Big)^T,\nonumber\\
&&\ket{2,k}=\Big(0,-\frac{S_y}{S_{xy}},\frac{S_x}{S_{xy}}\Big)^T,\nonumber\\
&&\ket{3,k}=\frac{1}{\sqrt{2}}
\Big(i,\frac{S_x}{S_{xy}},\frac{S_y}{S_{xy}}\Big)^T\nonumber
\ee
with $S_{xy}=\sqrt{S_x^2+S_y^2}$.

To find the collective operator which rotates among the four states $(\pm 1, \pm 1)$, let us first find the operator which rotates among the fermion states of these four collective states. Then the collective operator we are looking for is simply the collective (but uncondensed) part of such an operator formed from the fermions. Let us use subscript $a$, $b$, $c$, $d$ to label order parameter $(1,1)$, $(-1,-1)$, $(-1,1)$, $(1,-1)$. Then we can find a unitary operator which rotates among the four states.
\be
U_{ad}&=&\ket{1}_d\bra{1}_a+\ket{2}_d\bra{2}_a+\ket{3}_d\bra{3}_a\nonumber\\
&=&\left(\begin{array}{ccc}
1 & 0 & 0\\
0 & 1 & 2\Omega s_xc_y/s_{xy}^2\\
0 & -2\Omega s_xc_y/s_{xy}^2 & 1
\end{array}\right)+o(L)\nonumber
\ee
with $s_{xy}=\sqrt{s_x^2+s_y^2}$. Similarly, we have
\be
U_{ac}&=&\ket{1}_c\bra{1}_a+\ket{2}_c\bra{2}_a+\ket{3}_c\bra{3}_a\nonumber\\
&=&\left(\begin{array}{ccc}
1 & 0 & 0\\
0 & 1 & -2\Omega s_yc_x/s_{xy}^2\\
0 & 2\Omega s_yc_x/s_{xy}^2 & 1
\end{array}\right)+o(L)\nonumber
\ee
We can now introduce the generator of the above transformation
\be
&&g^f_{x,k}=2i\frac{s_yc_x}{E_0^2}
(p_{x,k}^{\dagger}p_{y,k}-p_{y,k}^{\dagger}p_{x,k}),\\
&&g^f_{y,k}=-2i\frac{s_xc_y}{E_0^2}
(p_{x,k}^{\dagger}p_{y,k}-p_{y,k}^{\dagger}p_{x,k})
\ee
In terms of the generators, we have
\be
U_{ac}=1+i\Omega g^f_{x,k},\quad U_{cb}=1+i\Omega g^f_{y,k},\nonumber\\
U_{bd}=1-i\Omega g^f_{x,k},\quad U_{da}=1-i\Omega g^f_{y,k}\nonumber
\ee
It follows that the fermion Hamiltonian,  $\sum_{\bf k} g^f_{x,k} + g^f_{y,k}$ serves to mix any of the fermion states corresponding to the collective states $(\pm 1, \pm 1)$ with a state rotating ${\bf \Omega}$ by $\pm \pi/2$. The current corresponding to this term runs within the unit-cell from one oxygen to the next in the clockwise direction as shown in Fig. (\ref{rot-op}) and the corresponding current in the anti-clockwise direction. It has already been shown \cite{aji-shekhter-varma-pr} that such a term is generated by the nearest neighbor interactions. It then follows that a collective current state of the same symmetry also exists which serves as a kinetic energy term mixing the collective configurations.
In the basis of the collective states $|\pm1, \pm1\rangle$, this has all the transformation properties and commutation rules of the operator $(S^1+T^1+S^2+T^2)$.

\section{Need for a 4-dimensional representation of the operator ${\bf S}$}
\label{4d}

As discussed the gauge transformation is  given by the rotation matrix
\be
U(\theta)=e^{i\theta_1S^3/2}e^{i\theta_2T^3/2}e^{i\theta_3K^{33}/2}
\ee
The classical AT model is invariant under this transformation. Under this rotation, the quantum term $S^1$, $T^1$ and $K^{11}$ will be transformed as follows
\be
&&U(\theta)S^1U(\theta)^{\dagger}=\cos\theta_3(\cos\theta_1S^1
-\sin\theta_1S^2)\nonumber\\
&&\qquad-\sin\theta_3(\sin\theta_1K^{13}+\cos\theta_1K^{23})\\
&&U(\theta)T^1U(\theta)^{\dagger}=\cos\theta_3(\cos\theta_2T^1
-\sin\theta_2T^2)\nonumber\\
&&\qquad-\sin\theta_3(\sin\theta_2K^{31}+\cos\theta_2K^{32})\\
&&U(\theta)K^{11}U(\theta)^{\dagger}=\cos\theta_1\cos\theta_2K^{11}
+\sin\theta_1\sin\theta_2K^{22}\nonumber\\
&&\qquad-(\cos\theta_1\sin\theta_2K^{12}+\sin\theta_1\cos\theta_2K^{21})
\ee
The quantum terms related by this transformation are equivalent to each other. In the mean time, the mean field ground state is also transformed to a new form. In the loop current state basis, the original ground state can be written as $\ket{G}=(c^2,sc,sc,s^2)^T$. After the transformation, we have $\ket{G'}=U(\theta)\ket{G}$ which is a superposition of the 4 states with complex number coefficients. Thus after the transformation, $\ep{S^y}$ is nonzero and depend on the gauge parameter $\theta_{1,2,3}$.

Since $\vM(\vq)$ as a physical observable is gauge invariant, it should not depend on the gauge parameter $\theta_{1,2,3}$. This gauge dependence should disappear if we make a corresponding unitary transformation on the local spin operator $S^{1,2,3}$. But now we meet a immediate difficulty. The transformation $U(\theta)$ is a 4 by 4 matrix simply given by
\be
U(\theta)=\mbox{diag}\{e^{i\phi_1},e^{i\phi_2},e^{i\phi_3},e^{i\phi_4}\}
\ee
with $\phi_1=\frac{\theta_1+\theta_2+\theta_3}{2}$, $\phi_2=\frac{\theta_1-\theta_2-\theta_3}{2}$,
$\phi_3=\frac{-\theta_1+\theta_2-\theta_3}{2}$ and
$\phi_4=\frac{-\theta_1-\theta_2+\theta_3}{2}$. Translating this transformation to
the basis of local spin 1 state, we find that the corresponding transformation is
\be
V(\theta)=\mbox{diag}\{e^{i\phi_1},e^{i\phi_2}+e^{i\phi_3},e^{i\phi_4}\}
\ee
Clearly, on can see that $V(\theta)$ is not a unitary transformation. Therefore we cannot use $V(\theta)$ to cancel the gauge dependent.

This difficulty arise because there are 4 states in the loop current state basis but there is only 3 states in the local spin 1 state basis. In order to overcome this difficulty, we have introduced a 4 dimensional representation of the spin 1 states given in the paper.

\section{Approximate Calculations of Form factor}
\label{formfac}

We first consider the form factor for the $L_z$. Since $L_z$ only has diagonal matrix elements, we do not need to use the rotation operator yet. The form factor is determined by the current density around the O-Cu-O triangle. To simplify the calculation, we approximate the triangle shape by a circle shape current density and also assume the density is a Gaussian distribution around the circle. Then, the wave-function of the local spin states in terms of cylindrical coordinates $\vr=(r\cos\phi,r\sin\phi,z)$ are given by
\be
&&\psi_{a,\pm1}=\frac{1}{\sqrt{2\pi} w}e^{-\frac{(r-r_s)^2}{4w^2}}
e^{-\frac{z^2}{4w^2}} e^{\pm i\phi},\\
&&\psi_{a,0}=\frac{1}{\sqrt{2\pi} w}e^{-\frac{(r-r_s)^2}{4w^2}}
e^{-\frac{z^2}{4w^2}}
\ee
Here we assume the radius of the circle is $r_s$ and centered around $R_a$ and $w$ specify the width of these circles. Now we can verify that $\psi_{a,1}$ gives a torus shape current density.
\be
\vj(\vr)&=&-\frac{i}{2}(\psi_{a,1}^{\dagger}
\nabla\psi_{a,1}-\psi_{a,1}\nabla\psi_{a,1}^{\dagger})\nonumber\\
&=&\frac{1}{2\pi w^2}e^{-\frac{(r-r_0)^2}{2w^2}}
e^{-\frac{z^2}{2w^2}}\hat{\bm{\phi}}
\ee
Similarly, $\psi_{\ds}$ gives a torus shape current density with opposite direction. $\psi_{0}$ is real function, so the current is zero. These 3 states correspond to $\ket{\pm1}$ and $\ket{0}$ point like angular momentum states. If we take $w\to\infty$ limit, we find a current circle $\vj(\vr)=\delta(r-r_0)\delta(z)\hat{\bm{\phi}}$.

We still have $L_z=\psi^{\dagger}L_z\psi$, where $L_z=-i\frac{\p}{\p\phi}$. The $\phi$ dependent part will give the same matrix elements as before. Now we also have an extra $r$ and $z$ dependent part as follows
\be
F_{z,a}(\vr)=\frac{1}{2\pi w^2}\exp\Big[-\frac{(r_a-r_s)^2}{2w^2}-\frac{z^2}{2w^2}\Big]
\label{Fza}
\ee
Here $r_a=\sqrt{(x-R_{a,x})^2+(y-R_{a,y})^2}$ with $a=1,2,3,4$.
Then the form factor is the Fourier transformation of the above function.
If the current width is narrow, we have
\be
&&F(\vk)=\int_{\mbox{cell}} F(\vr)e^{-i\vk\cdot\vr}d^3r
\approx f(k_r)e^{-\frac{w^2k_z^2}{2}}e^{-i\vk\cdot\vR_a}\nonumber\\
&&\mbox{with}\qquad f(k_r)=\frac{1}{\sqrt{2\pi}w}\int_0^{\infty}e^{-\frac{(r-r_s)^2}{2w^2}}
J_0(k_rr)rdr\nonumber
\ee
Using this result in Eq. (\ref{Mz}), we find
\be
L_z(\vk)=\cos\theta\Big[F_{z,1}(\vk)e^{-i\vk\cdot\vR_1}-F_{z,1}(\vk)e^{-i\vk\cdot\vR_3}\Big]
\ee

The form factor of $L_{t}$ is more complicated, since the rotation operator has its own coordinate dependence. As explained before, the rotation operator is currents flows around the four oxygens in the unit cell. Here for simplicity, we approximate the current around the oxygens as a big circle with radius $r_0$ centered at $\vR_i^0$. We have $L_{t}=\psi^{\dagger} L_{t}\psi$. Here $L_{t}$ will annihilate or create the $e^{i\phi}$ factor. The $\phi$ dependent part will give the same matrix elements as before. For the local spin state at $\vR_a$, we also have an extra $r$ and $z$ dependent part as follows
\be
F_{x,a}(\vr)&=&\frac{1}{2\pi w^2}\exp\Big[-\frac{1}{2w^2}(r-r_0)^2
-\frac{z^2}{2w^2}\Big]\nonumber\\
& &\times\exp\Big[-\frac{1}{2w^2}(r_a-r_s)^2
-\frac{z^2}{2w^2}\Big]
\label{Fxa}
\ee
Here $r=\sqrt{x^2+y^2}$. It is easy to transform the $z$ dependent part to momentum space. The $x,y$ dependent part has to be computed by numerics. Including the form factors in Eq. (\ref{Mx}), the x-component of moment is
\be
L_t(\vk)&=&\sin\theta\Big[F_{x,1}(\vk)e^{-i\vk\cdot\vR_1}
+F_{x,2}(\vk)e^{-i\vk\cdot\vR_2}\nonumber\\
& &-F_{x,3}(\vk)e^{-i\vk\cdot\vR_3}-F_{x,4}(\vk)e^{-i\vk\cdot\vR_4}\Big]
\ee
If the current width is quite narrow $w\ll a$, here $a$ is lattice constant of xy-plane,  then Eq. (\ref{Fxa}) can be approximated by to two Gaussian peaks located at $\vR_{a}^1$ and $\vR_a^2$ where $\vR_a^1$ and $\vR_a^2$ are the two intersection points of the two circles. Thus we have
\be
F_{x,a}(\vr)\approx\frac{1}{2\pi w^2}\Big[e^{-\frac{(\vr-\vR_a^1)^2}{2w^2}}
+e^{-\frac{(\vr-\vR_a^2)^2}{2w^2}}\Big]e^{-\frac{z^2}{w^2}}
\ee
Transferring to momentum space, we find the form factor as
\be
F_{x,a}(\vk)=\sqrt{\pi}w e^{-\frac{w^2(k_x^2+k_y^2)}{2}}
\Big(e^{i\vk\cdot\vR_a^1}+e^{i\vk\cdot\vR_a^1}\Big)e^{-\frac{w^2k_z^2}{4}}
\label{Fxak}
\ee
with $a=1,2,3,4$.

So far we only considered the ``spin" kind contribution to the moments. Similarly, we can also compute the form factor for the ``rotation" kind contribution. Follow the same line of arguments,  the $\phi$ dependent part will give the same matrix elements as before. For the the matrix element $\bra{\psi_a}L_{t}\ket{\psi_b}$, there is an extra $r$ and $z$ dependent part as follows
\begin{widetext}
\be
&&F_{x,ab}(\vr)=\frac{1}{2\pi w^2}\exp\Big[-\frac{1}{2w^2}(r-r_0)^2-\frac{z^2}{2w^2}\Big]
\exp\Big[-\frac{1}{4w^2}(r_a-r_s)^2-\frac{z^2}{4w^2}\Big]\nonumber\\
&&\qquad\qquad\times\exp\Big[-\frac{1}{4w^2}(r_b-r_s)^2-\frac{z^2}{4w^2}\Big]
\label{Fxab}
\ee
Similarly, or the the matrix element $\bra{\psi_a}L_{z}\ket{\psi_b}$, there is an extra $r$ and $z$ dependent part as follows
\be
F_{z,ab}(\vr)=\frac{1}{2\pi w^2}\exp\Big[-\frac{1}{4w^2}(r_a-r_s)^2-\frac{z^2}{4w^2}\Big]
\exp\Big[-\frac{1}{4w^2}(r_b-r_s)^2-\frac{z^2}{4w^2}\Big]
\label{Fzab}
\ee
And the form factor is the above function transform to momentum space. Therefore, for transfer momentum $\vk=\frac{2\pi}{a}(0,1,0)$, we find the effective moment due to the ``rotation" kind contribution is
\be
\vL_{\mbox{eff}}=-ik_yC\Big(F_{x,12}(\vk)(1+\sin^2\theta),\,0,\,F_{z,12}(\vk)\sin2\theta\Big)
\ee

These results have been used in the paper to calculate the form-factor and the tilt angles and compared with experiments.
\end{widetext}

\end{document}